\ifx\documentclass\undefined  % LaTeX2e?
  \documentstyle[11pt,a4]{article}
\else
  \documentclass[12pt,a4paper]{article}
\fi

\begin{document}
\begin{center}
{\bf Persistent X-Ray Photoconductivity and Percolation of Metallic Clusters
in Charge-Ordered Manganites}

\vspace{.9in}

D. Casa, V.  Kiryukhin$^*$, O.A. Saleh, and B. Keimer$^\dagger$ \\

\vspace{.05in}

Dept. of Physics, Princeton University, Princeton, NJ 08544, USA, and\\
$^\dagger$  Max-Planck-Institut f\"ur Festk\"orperforschung, D-70569 Stuttgart, 
Germany

\vspace{.3in}

J.P. Hill \\

\vspace{.05in}

Dept. of Physics, Brookhaven National Laboratory, Upton, NY 11973, USA 

\vspace{.3in}

Y. Tomioka and Y. Tokura$^{\dagger\dagger}$ \\

\vspace{.05in}

Joint Research Center for Atom Technology (JRCAT), Tsukuba, Ibaraki 305, Japan,
and\\
$^{\dagger\dagger}$ Dept. of Applied Physics, University of Tokyo, Tokyo 113, 
Japan

\vspace{.6in}

ABSTRACT
\end{center}

\vspace{.1in}

\noindent Charge-ordered manganites of composition $\rm Pr_{1-x}(Ca_{1-
y}Sr_{y})_{x}MnO_3$  
exhibit persistent photoconductivity upon exposure to x-rays. This is not always
accompanied by a significant increase in the {\it number} of conduction 
electrons as predicted by conventional models of persistent photoconductivity.
An analysis of the x-ray diffraction patterns and current-voltage 
characteristics shows 
that x-ray illumination results in a microscopically phase separated state in 
which 
charge-ordered insulating regions provide barriers against charge transport 
between 
metallic clusters. The dominant effect of x-ray illumination is to enhance the 
electron 
{\it mobility} by lowering or removing these barriers. A mechanism based on 
magnetic 
degrees of freedom is proposed.

\vspace{.05in}


\vspace{.3in}

\noindent $^*$ Present address: Dept. of Physics, 13-2154, MIT, Cambridge, MA 
02139

\clearpage

The manganite perovskite $\rm Pr_{0.7} Ca_{0.3} MnO_3$ has recently been shown 
to 
undergo an unusual insulator-metal transition when it is exposed to an x-ray 
beam \cite{kiryukhin}. Without x-rays, the material is 
semiconducting (with an average valence of +3.3 distributed uniformly over the 
Mn
sites) at high temperatures, and charge-ordered insulating (with a static 
superlattice
of $\rm Mn^{3+}$ and $\rm Mn^{4+}$ ions) below $\sim$ 200K. Below $\sim$ 40K, 
x-rays convert the insulating state to a metallic state which persists when the 
x-ray
beam is switched off, but is removed on thermal cycling \cite{kiryukhin}.
Interestingly, an equilibrium insulator-metal transition can also be induced by 
an external magnetic field \cite{tomioka}, but not by irradiation with 
visible light. The material is, however, susceptible to dielectric breakdown at 
comparatively low electric fields \cite{asamitsu} that can be triggered by laser 
light
\cite{miyano}. A microscopic explanation for this unusual behavior has thus far 
not
been given. In this paper, we take the first steps in this direction by 
proposing a new
mechanism for persistent photoconductivity (PPC), based on a quantitative 
analysis of new
field and illumination dependent transport and diffraction data. Whereas a 
change in the {\it number}
of conduction electrons is thought to be at the origin of PPC 
in conventional semiconductors, our data indicate that the electron {\it 
mobility} is the decisive factor in
the manganites.  The mobility is controlled mostly by magnetic degrees of
freedom, that is, by the strong coupling of the conduction electrons (in the Mn 
$e_g$ orbitals) 
to local spins (in the Mn $t_{2g}$ orbitals) via Hund's rule. 

The measurements were performed in a vertical-field 
superconducting magnet mounted on a two-circle
goniometer at beamline X22B (photon energy 8 keV, flux $\rm 5 \times 10^{10}/
sec$) at the National Synchrotron Light Source. An ion chamber
was used to monitor the incident x-ray beam. The x-ray fluence (that is, the
cumulative number of photons incident per $\rm mm^2$ of sample surface) was
estimated from the monitor counts. The diffracted beam
was detected by a scintillation detector. The samples were single crystals of 
$\rm Pr_{1-x} (Ca_{1-y} Sr_y)_x MnO_3$ synthesized by a floating zone
method described previously \cite{tomioka}. 
The crystal surface was polished with diamond paste to better than 1 $\mu$m.
Two 4000 $\rm \AA$ thick Au electrical contacts about 2 mm apart were evaporated
on a 200 $\rm \AA$ thick Cr buffer layer and 
annealed at 500$^\circ$C for 10 hours. The contact resistance was less than 
1$\Omega$. The x-ray beam illuminated the region between
the contact pads.

This experimental setup allows simultaneous measurements of the electrical
conductivity and of the intensities of superstructure reflections characteristic 
of 
the cooperative Jahn-Teller distortion in the insulating state. These 
reflections occur at 
commensurate reciprocal lattice vectors (H, K/2, L), K odd, indexed
on an orthorhombic lattice \cite{kiryukhin,jirak,yoshizawa}. For reference, we 
have reproduced the
central observations \cite{kiryukhin} on $\rm Pr_{0.7} Ca_{0.3} MnO_3$ in Fig. 
1a.
The x-ray penetration depth is about 2 $\mu$m, and the enhancement of the
electrical conductivity by several orders of magnitude shows that a thin 
metallic 
film is created on an insulating background when the material is exposed to x-
rays 
(inset in Fig. 1). These metallic regions
replace charge-ordered insulating regions, as indicated by the x-ray induced 
reduction
of the superlattice peak intensity.

We have carried out a more comprehensive 
study of $\rm Pr_{0.7} Ca_{0.3} MnO_3$ as well as analogous measurements
on $\rm Pr_{0.65} Ca_{0.245} Sr_{0.105} MnO_3$ and $\rm Pr_{1-x} Ca_x MnO_3$
with x=0.4 and x=0.5, all of which exhibit charge ordering with the same wave 
vector
as $\rm Pr_{0.7} Ca_{0.3} MnO_3$. The average Mn valence for x=0.5 is 3.5, and a
$\rm Mn^{3+}$-$\rm Mn^{4+}$ superlattice with a doubled unit cell can form 
with a minimum number of defects. For
x=0.4 and 0.3, additional electrons have to be accommodated in defects or domain
boundaries, and the charge-ordered lattice is less robust. This is reflected in 
the
magnetic phase diagrams \cite{tomioka} which show that an external magnetic 
field
induces a strongly first order insulator-metal transition associated with a 
large
hysteresis region; progressively lower magnetic fields are required to induce 
this
transition as x decreases. For the Sr-substituted compound, the critical field 
is even
lower, and the metallic state reenters at low temperature even without 
application
of a magnetic field. Some magnetic phase diagrams are reproduced in the insets 
of
Fig. 1. As the charge-ordered lattice is antiferromagnetically ordered at low
temperatures, defects and domain boundaries are expected to give rise to spin
disorder. 

While we observed an x-ray induced conductivity enhancement of the same order of
magnitude as in $\rm Pr_{0.7} Ca_{0.3} MnO_3$ in all of these compounds 
throughout 
most of the hysteresis region in
the H-T phase diagrams (Fig. 1), a depression of the superlattice reflection was 
only observed in a narrower region close to the metallic side of the hysteresis 
region. 
An example is given in Fig. 1b for $\rm Pr_{0.65} Ca_{0.245} Sr_{0.105} MnO_3$
where the peak intensity is independent of illumination at the 1\% level. Note
that in this compound the photoeffect occurs at a higher temperature where the
intrinsic resistivity without x-rays ($\sim 10^3 \Omega$cm) is much smaller than 
in 
$\rm Pr_{0.7} Ca_{0.3} MnO_3$ at low temperatures ($>10^6 \Omega$cm). Since the
resistance of the 2 $\rm \mu m$ thick layer affected by the x-rays and the bulk 
of
the crystal that remains unaffected are measured 
in parallel, the total x-ray induced resistance drop is much lower than in the 
$\rm Pr_{0.7} Ca_{0.3} MnO_3$ experiment. The x-ray induced change in 
resistivity
within the volume penetrated by the x-rays is comparable, however. 

Closely similar observations were made in the other compounds, indicating that, 
despite the observations in $\rm Pr_{0.7} Ca_{0.3} MnO_3$, persistent x-ray 
photoconductivity does not, {\it in general}, require the destruction of a 
significant
amount of the charge-ordered phase. Rather, these new data suggest that 
substantial
x-ray induced phase conversion, out of the charge-ordered insulating phase, 
takes
place only in a region of the phase diagram in which the metallic state is
thermodynamically stable and the charge-ordered state is metastable. Since in 
$\rm Pr_{0.65} Ca_{0.245} Sr_{0.105} MnO_3$, under the conditions of Fig. 1b, at
most a very small amount of the charge-ordered phase is converted, the 
corresponding
increase in the number of conduction electrons is also small, certainly much 
smaller
than in $\rm Pr_{0.7} Ca_{0.3} MnO_3$ (Fig. 1a). Nevertheless, both materials 
show
comparable PPC. This observation is at odds with
conventional models of PPC in semiconductors that rely exclusively on a
photogenerated increase in the {\it number} of free charge carriers \cite{lang}.
(According to these models, the carriers are photoexcited out of impurity 
states, and
lattice relaxations prevent optical recapture.) By contrast, the decisive
factor in the manganites appears to be the {\it mobility} of the electrons.

In order to elucidate how the x-ray photoelectrons affect the mobility of the
charge carriers, we have carried out a systematic study of the current-voltage 
(I-V)
characteristics in $\rm Pr_{0.7} Ca_{0.3} MnO_3$. 
A synopsis of the evolution of the I-V curves with x-ray
illumination in zero field, and with magnetic field after a brief x-ray exposure 
at
T=5K, is given in Fig. 2. By repeating the magnetic field dependent measurements
without x-ray exposure, we verified
that for all curves of Fig. 2b the conductivity is dominated by the conducting 
film 
created by the initial brief x-ray illumination, with negligible contribution 
of the bulk of the crystal. Both sets of curves in Fig. 2 can thus be directly 
compared
and, remarkably, turn out to be closely similar. 

Whereas in both cases the transport is ohmic
after prolonged x-ray exposure or in a high magnetic field, indicating a 
continuous
metallic path between the contacts, the I-V characteristics are initially
highly nonlinear. There are different possible origins of nonohmic conductivity
in insulators, notably Joule heating, variable range hopping, and tunneling. The 
first
two are ruled out by the nearly temperature independent conductivity below T=10K
shown in Fig. 3. Further, detailed consideration of the variable range hopping 
scenario
\cite{pollak} reveals that the electric fields at which such effects are 
predicted to
become observable are several orders of magnitude larger than the weak fields
applied here.

This leaves tunneling between x-ray induced isolated metallic islands as the 
only
viable model. A surprisingly good fit to most curves can be obtained
under the assumption that the transport is dominated by a small number of tunnel
junctions in series, each of which is described by the well known Simmons model
\cite{simmons} with physically resonable parameters (height of the insulating
barrier $\sim 1-2$eV, thickness $\rm \sim 10-30\AA$). These parameters
are also consistent with those determined on a single
trilayer manganite tunnel junction at comparable electric fields \cite{sun}. A 
typical result of such a comparison is shown in the inset of Fig. 3. A 
comprehensive analysis of all I-V curves, including a possible special role of 
insulating barriers near the contacts, is beyond the
scope of this paper. 

The picture that emerges from this analysis is one in which a small number of
insulating barriers act as ``bottlenecks'' for charge transport 
between the ferromagnetic metallic droplets. (Especially in narrow barriers, 
significant disorder may be present in both the charge ordering pattern and the 
antiferromagnetic 
spin alignment \cite{yoshizawa}.) This explains why the x-ray induced 
enhancement of the conductivity
is comparable in $\rm Pr_{0.7} Ca_{0.3} MnO_3$ and $\rm Pr_{0.65} Ca_{0.245} 
Sr_{0.105} MnO_3$
despite the vastly different number of conduction electrons added through phase 
conversion:
The decisive step in both cases is an x-ray induced enhancement of the electron 
mobility by lowering or removing
the insulating barriers. The parallel evolution of the I-V curves in Figs. 2a 
and b suggests
that this proceeds in a very similar fashion upon x-ray illumination and upon 
application of a magnetic field. The effect of a magnetic field on these 
materials is much better understood 
than the effect of x-rays: In the generally accepted ``double exchange'' model, 
the field 
aligns the local $t_{2g}$ spins ferromagnetically and facilitates hopping of the
conduction electrons between adjacent Mn sites. This is believed to be the 
primary
origin of the ``colossal magnetoresistance'' in the manganites. In the present 
case, 
the mobility through the barriers is enhanced if application of the field leads 
to
canting of the local spins in the barriers. For a large enough field,
some of the barriers may be entirely converted to the ferromagnetic phase
so that adjacent metallic clusters coalesce. Fig. 2 suggests that these two 
effects (or
a combination thereof) are also elicited by
the x-rays. In $\rm Pr_{0.65} Ca_{0.245} Sr_{0.105} MnO_3$, phase conversion is 
negligible
(Fig. 1b) and the dominant effect has to be a modification of the spin alignment 
in the insulating
barriers by the x-rays \cite{note}. A possible mechanism involves penetration of 
the barriers
(which are insulating to conduction electrons in the clusters) by hot x-ray 
photoelectrons which 
then actuate the double-exchange mechanism. 

The importance of magnetic degrees of freedom in controlling the conductivity 
across
the insulating barriers is underscored by I-V curves taken near the percolation 
threshold of the
metallic clusters. Here, the I-V curves become unstable
and exhibit frequent sudden jumps even for extremely low injected currents (Fig. 
4). Interestingly,
in the intermediate regime the behavior of the transport characteristics
under x-ray exposure (Fig. 4a) and in a magnetic field (Fig. 4b) is 
systematically
different. In the former case, jumps into states with higher and lower 
conductivity 
are approximately equally likely, while in the
latter case, jumps occur exclusively into more highly conducting states. 

This observation fits naturally 
into the above scenario in which the conductivity near percolation is 
limited by tunneling through very narrow residual insulating regions with 
Mn atoms in metastable spin configurations. Local heating 
by a minute transport current allows these spins to explore different 
configurations.
In the presence of an aligning field (Fig. 4b; the curves were taken with the 
magnetic 
field on), the final configurations are generally characterized by enhanced
ferromagnetic correlation more conducive to charge transport. 
In the absence of an aligning field (Fig. 4a; the curves were taken with the x-
ray beam
off), there is no preferential spin alignment after the heating event, and the
conductivity is equally likely to be reduced or enhanced. This demonstrates 
directly that 
nonequilibrium electrons can affect the mobility through the barriers by 
affecting their spin
configuration.

In summary, we have identified a novel mechanism of PPC
which is operative only for ``hot'' photoelectrons generated by high frequency
radiation. Hot electrons appear to move an appreciable distance
into the charge-ordered regions and enhance the electron mobility. The mechanism 
may involve a nonequilibrium carrier population near the top of the Mn $e_g$ 
band that 
interacts with the local spins via double exchange. As 
electrons in the electric breakdown regime are also characterized by an 
effective temperature much higher 
than the crystalline environment, hot electron transport in these magnetically 
correlated
materials provides a unified microscopic framework for both the photoinduced and 
the electric
field induced insulator-metal transitions.

\vspace{.1in}

\noindent {\bf Acknowledgments.} This work was supported by the National 
Science Foundation under grant No. DMR-9701991 (B.K.), by the Packard and Sloan
Foundations (B.K.), by the US-DOE under contract No. DE-AC02-98CH10886 (J.P.H.), 
and by NEDO 
and Grants-In-Aid from the Ministry of Education, Japan (Y.T.).

\clearpage

\section*{Figure Captions}
\begin{enumerate}

\item Electrical resistance and intensity of the (2, 1.5, 0) superlattice 
reflection
characteristic of charge ordering as a function of x-ray exposure for 
(a) $\rm Pr_{0.7} Ca_{0.3} MnO_3$ at T=5K, and (b) $\rm Pr_{0.65} Ca_{0.245} 
Sr_{0.105} MnO_3$ at T=100K. The insets in the lower panels show the magnetic 
phase diagrams \cite{tomioka}. Significant hysteresis is observed in the shaded 
regions, and the dashed line
represents the estimated thermodynamic phase boundaries following Ref. 
\cite{tokura}. The inset
in the upper right panel illustrates the experimental geometry.

\item Current-voltage characteristics of $\rm Pr_{0.7} Ca_{0.3} MnO_3$ for (a)
various x-ray exposures in zero field and (b) magnetic fields after a brief 
x-ray exposure ($\rm 4.4 \times 10^{12} \; photons/mm^2$), at T=5K. 
The curves in (a) are labeled
by the incident x-ray fluence in units of $10^{13}$ photons per $\rm mm^2$ 
incident on the sample.
Those in (b) are labeled by the magnetic field in units of Tesla.

\item Temperature dependence of the current at a fixed voltage of 4V after a 
brief
x-ray exposure ($\rm 5.5 \times 10^{12}\; photons/mm^2$). 
The data in the inset were measured at H=2T after an exposure of 
$\rm 4.4 \times 10^{12} \; photons/mm^2$, the line is the Simmons tunneling 
expression \cite{simmons} for two series junctions
with barrier height 1.5 eV and thickness 15$\rm \AA$.

\item Typical current-voltage characteristics close to the percolation threshold 
of
(a) x-ray generated and (b) magnetic field generated metallic clusters. The 
curves in
(a) were taken sequentially at the same x-ray exposure ($\rm 1.3\times10^{14} \;
photons/mm^2$), with
the x-ray beam off. The curves in (b) were taken sequentially at a field of 
3.75T, with
the field on.

\end{enumerate}

\begin{thebibliography}{99}
\bibitem{note} In $\rm Pr_{0.65} Ca_{0.245} Sr_{0.105} MnO_3$, some charge 
delocalization
may take place in the barriers without a noticeable 
effect on the superlattice peak intensity (Fig. 1b), if this involves mostly Mn 
ions with local (incoherent or partially coherent) Jahn-Teller distortions.
\end{thebibliography}
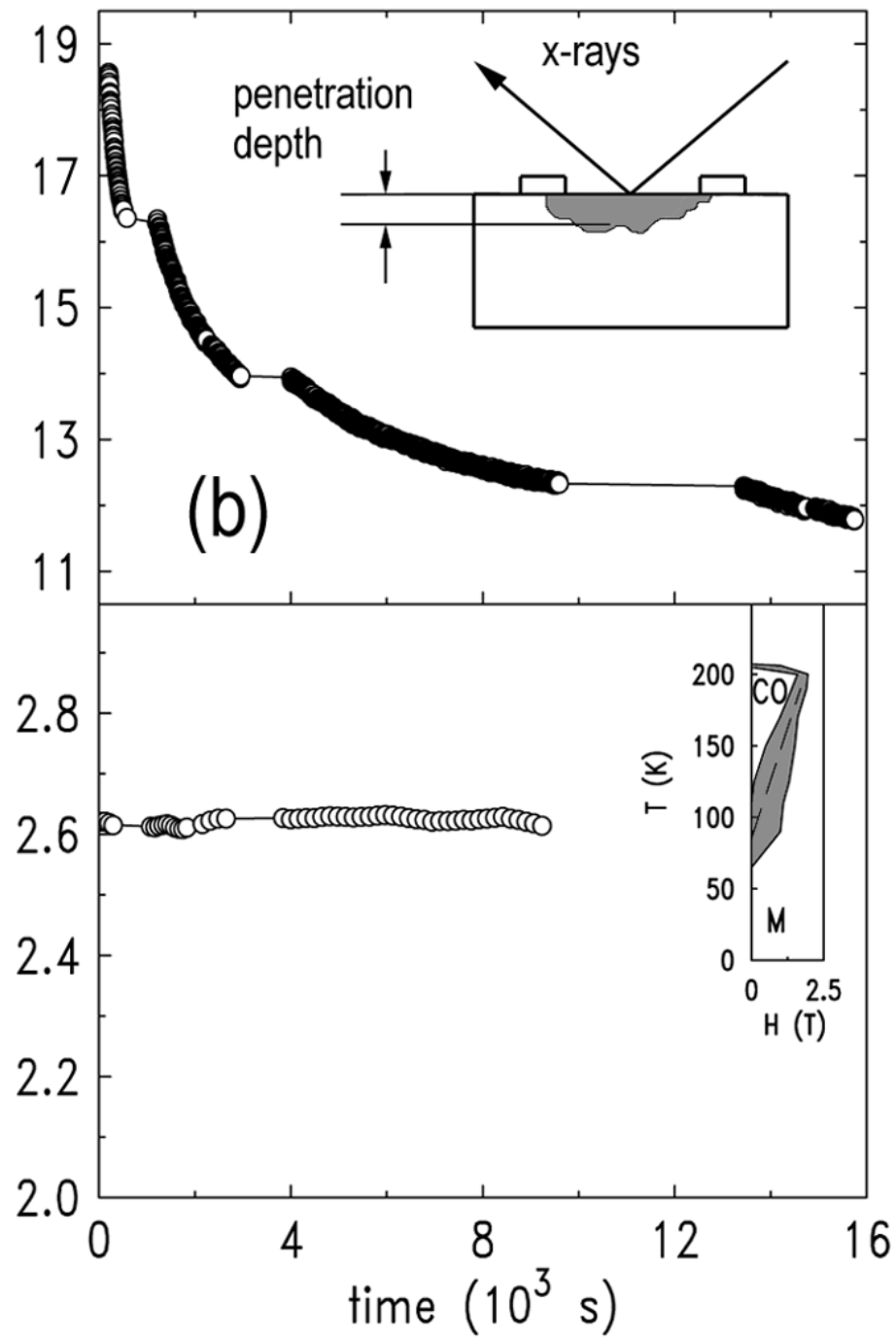
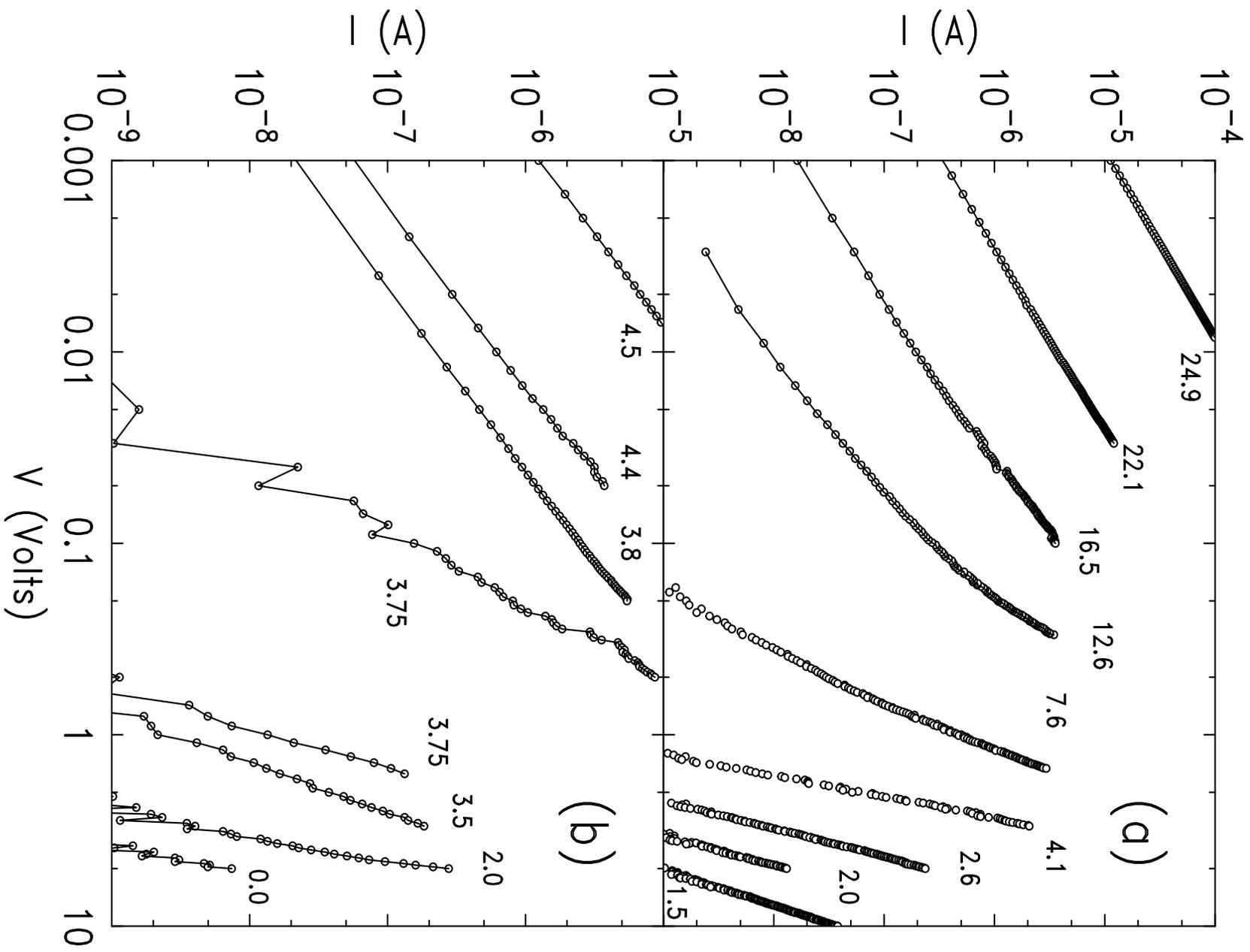
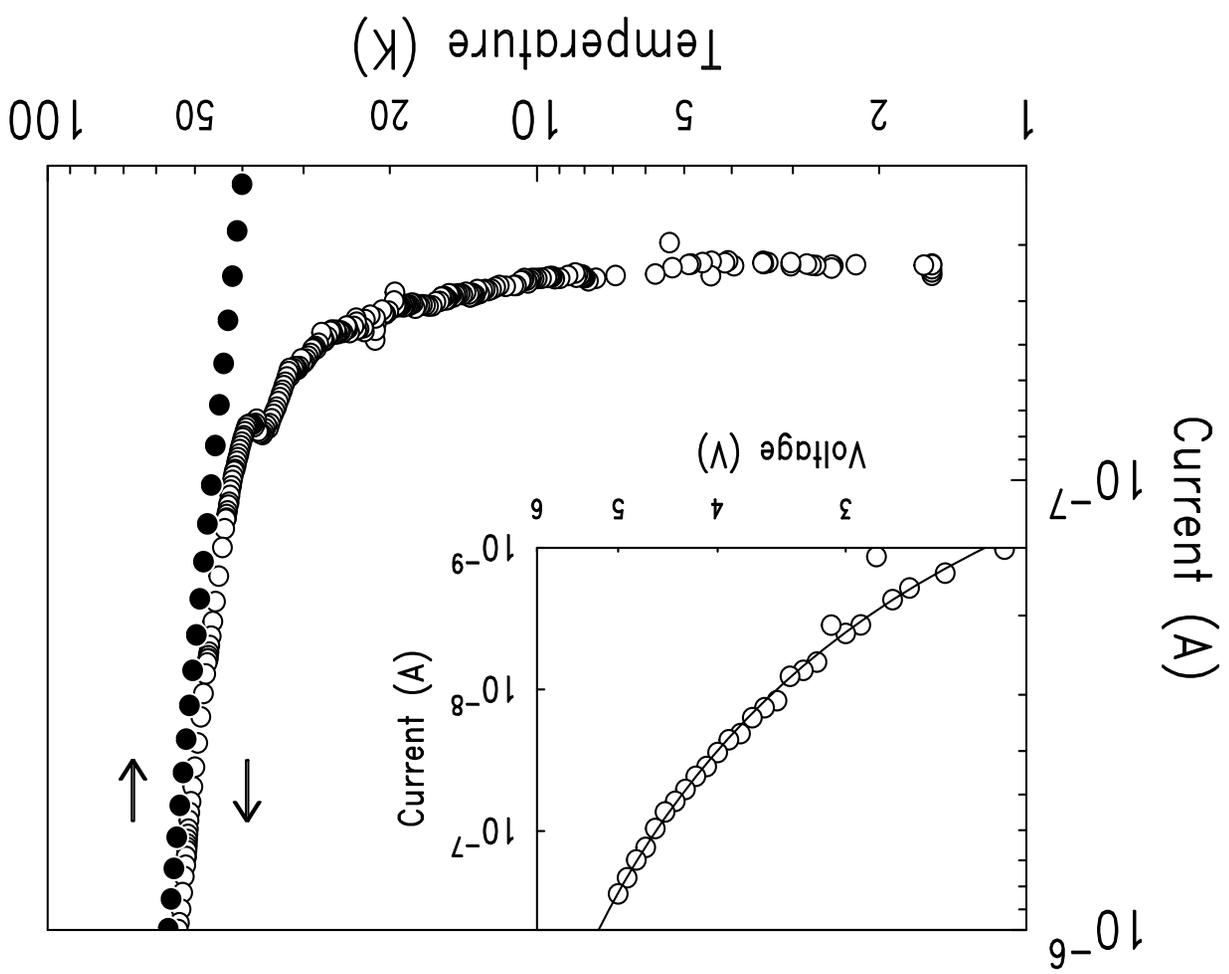
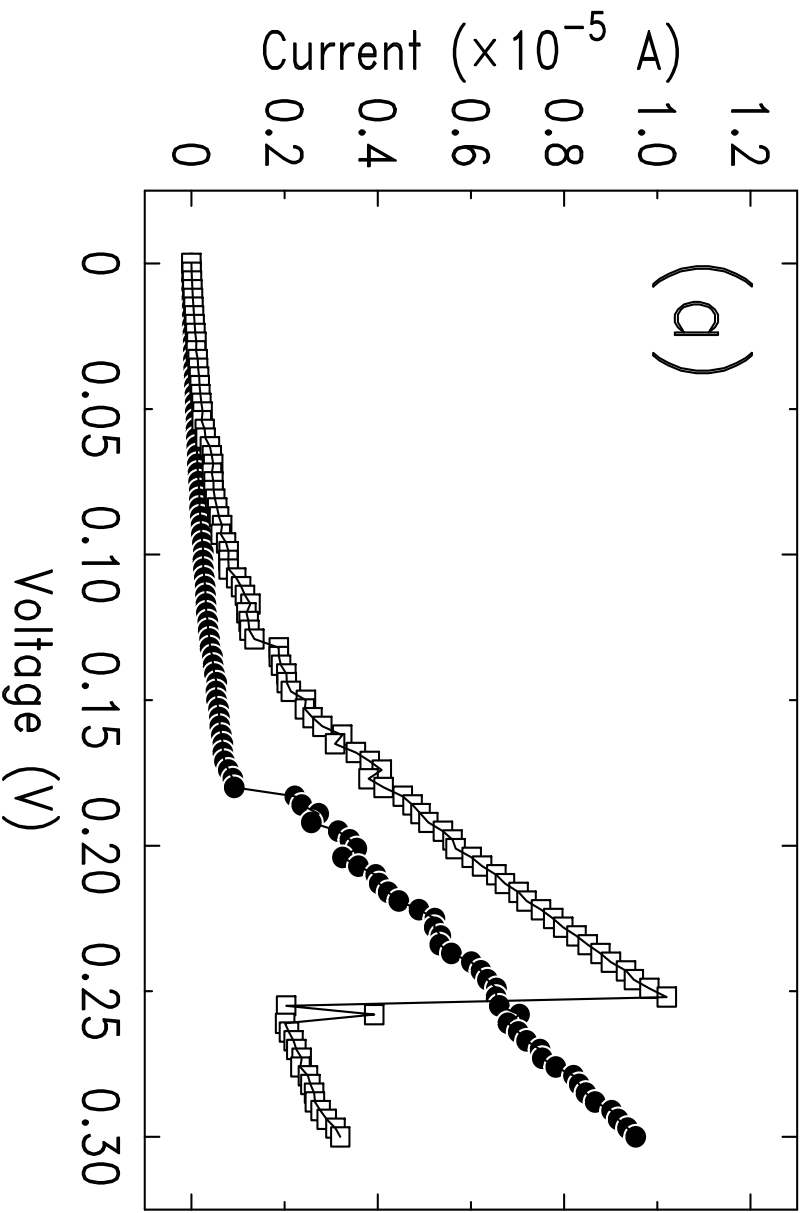
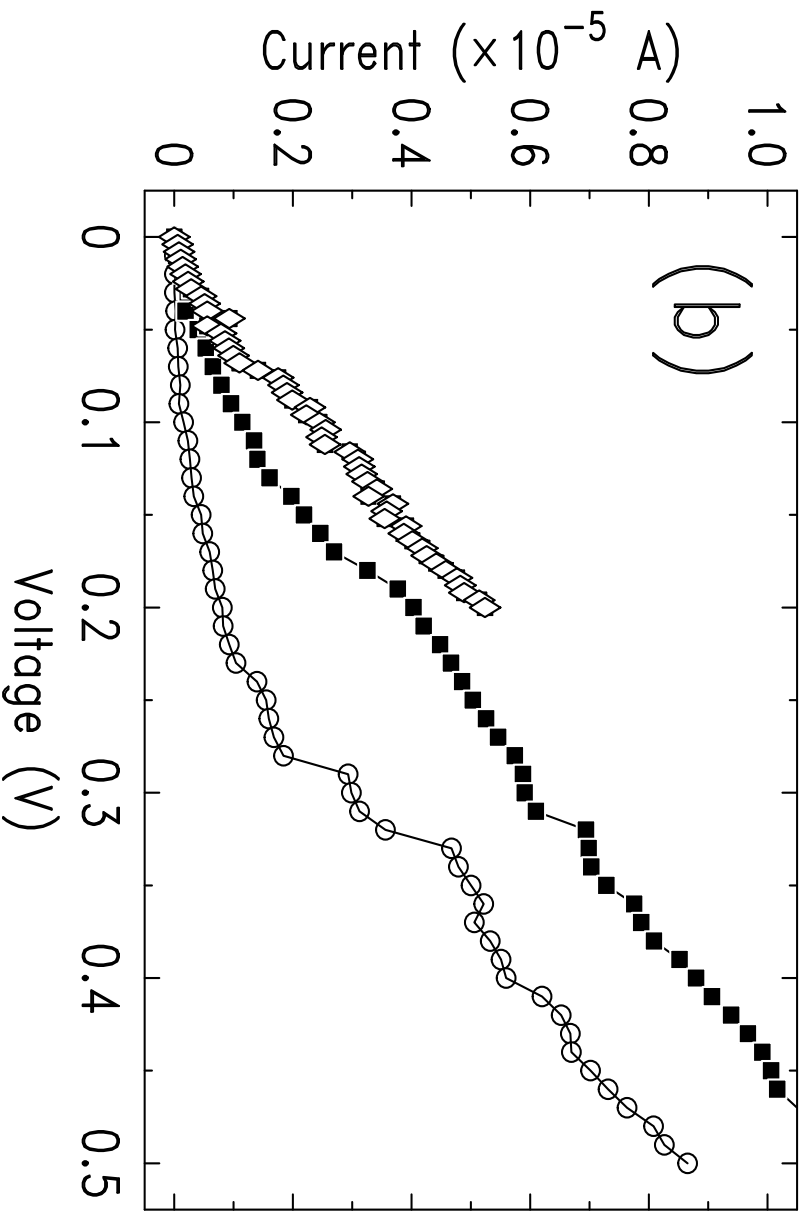
\end{document}